\documentclass{article}

\usepackage{arxiv}
\usepackage{xcolor}
\usepackage{authblk}
\usepackage{amssymb}
\usepackage{amsmath}
\usepackage{multicol}
\usepackage{listings}

\definecolor{CodeBackground}{rgb}{0.98,0.98,0.98}
\definecolor{TypeBackground}{rgb}{0.95,0.95,0.95}
\definecolor{DarkGreen}{rgb}{0.0,0.5,0.0}
\definecolor{Brown}{cmyk}{0,0.81,1,0.60}
\definecolor{OliveGreen}{cmyk}{0.64,0,0.95,0.40}
\definecolor{CadetBlue}{cmyk}{0.62,0.57,0.23,0}
\definecolor{val-color}{rgb}{0.8, 0.47, 0.196}
\definecolor{event-color}{rgb}{0.433, 0.528, 0.145}
\definecolor{data-color}{rgb}{0.2433, 0.128, 0.345}

\lstdefinelanguage{sol}{
	language = Java,
	morekeywords={internal, returns, function, constructor, modifier, payable, modifier},
	deletekeywords={this},
	morekeywords=[2]{bool, int, int32, uint, uint256, bytes32, mapping, address, storage, enum, struct},
	morekeywords=[3]{contract},
	morekeywords=[4]{require, emit, event, using, pragma, library},
	keywordstyle=\color{blue},	
	keywordstyle=[2]{\color{data-color}},
	keywordstyle=[3]{\color{val-color}},
	keywordstyle=[4]{\color{event-color}},
	commentstyle=\color{gray},
	sensitive=false, 
	morecomment=[l]{//}, 
	morecomment=[s]{/*}{*/}, 
	morestring=[b]" 
}

\lstdefinelanguage{ov}{
	language = Java,
	morekeywords={constructor, payable},
	deletekeywords={this},
	morekeywords=[2]{bool, int, int32, uint, uint256, bytes32, map, address, enum},
	morekeywords=[3]{o, oo, this, top, bot, atomic, object},
	morekeywords=[4]{require, emit, event},
	keywordstyle=\color{blue},	
	keywordstyle=[2]{\color{data-color}},
	keywordstyle=[3]{\color{val-color}},
	keywordstyle=[4]{\color{event-color}},
	commentstyle=\color{gray},
	sensitive=false, 
	morecomment=[l]{//}, 
	morecomment=[s]{/*}{*/}, 
	morestring=[b]" 
}

\lstdefinestyle{ov}
{
	language = Java,
	basicstyle = {\ttfamily \color{main-color}},
	otherkeywords = {object},
	keywordstyle = {\color{key-color}},
}

\newtheorem{thm}{Theorem}[section]
\newtheorem{lem}[thm]{Lemma}

\newcommand{\ov}{{\rm OV}}

\newcommand{\avid}{{\rm AVID}}

\newcommand{\keyword}[1]{\textsf{#1}}
\newcommand{\ruledef}[3]{\begin{tabular}{lr}\rulename{#1}\\ \quad $\frac{\begin{array}[c]{c}#2\end{array}}{\begin{array}[c]{c}#3\end{array}}$\end{tabular}}

\newcommand{\rulename}[1]{{\footnotesize\texttt{[\MakeUppercase{#1}]}}}

\newcommand{\methsyntax}{T\ m(\overline{T\ x})\ \Delta\ \{e\}}

\newcommand{\type}[1]{\Gamma\vdash#1}

\newcommand{\typeO}[1]{\Gamma\vdash_o#1}
\newcommand{\subtype}[2]{\vdash#1<#2}
\newcommand{\binding}[2]{\vdash#1<:#2}

\newcommand{\abstraction}[2]{\vdash#1\ll#2}
\newcommand{\abstractionM}[2]{\vdash\overline{#1\ll#2}}
\newcommand{\context}[1]{\Gamma\vdash#1}

\newcommand{\constraintM}[2]{\Gamma\vdash\overline{#1\ R\ #2}}

\newcommand{\insideeq}[2]{\Gamma\vdash#1\preceq#2}

\newcommand{\exprX}[3]{#1;\Delta\vdash#2:#3}
\newcommand{\exprXX}[4]{#1;#2\vdash#3:#4}
\newcommand{\exprE}[2]{\vdash#1:#2}
\newcommand{\exprEX}[3]{#1\vdash#2:#3}

\newcommand{\contract}[1]{\Gamma\vdash#1}
\newcommand{\subcontract}[2]{\Gamma;#1\vdash#2}

\newcommand{\subsetcon}[2]{\Gamma\vdash#1\sqsubseteq#2}

\newcommand{\modelV}{\mathcal{V}}
\newcommand{\modelI}{\mathcal{I}}
\newcommand{\modelVI}{\langle \mathcal{V},\mathcal{I} \rangle}
\newcommand{\modelVIX}{\langle \mathcal{V}',\mathcal{I}' \rangle}

\newcommand{\valcontract}[1]{$\langle${#1}$\rangle$}

\newcommand{\modelVIn}[1]{\langle \mathcal{V}_{#1},\mathcal{I}_{#1} \rangle}
\newcommand{\tran}[1]{$#1$}

\newcommand{\atomic}{\keyword{atomic}}
\newcommand{\valid}{\keyword{valid}}

\newcommand{\textcode}[1]{{\small\texttt{#1}}}

\newcommand{\reduce}[6]{\alpha;H;\Sigma;S;#1\longrightarrow#2;#3;#4;#5;#6}

\newcommand{\reduceX}[2]{#1\longrightarrow#2}
\newcommand{\heap}[0]{\vdash H}
\newcommand{\heapX}[1]{\vdash #1}

\newcommand{\validity}[0]{H;\Sigma\vdash \Delta}
\newcommand{\validityX}[3]{#1;#2\vdash #3}

\renewcommand{\and}{\hspace{.5cm}}

\title{\ov: Validity-based Optimistic smart contracts}

\author{Quan Nguyen \and Andre Cronje \and Michael Kong}

\affil{FANTOM}

\begin{document}
\maketitle

\begin{abstract}
	Smart contract (SC) platforms form blocks of transactions into a chain and execute them via user-defined smart contracts.
	In conventional platforms like Bitcoin and Ethereum, the transactions within a block are executed \emph{sequentially} by the miner and are then validated \emph{sequentially} by the validators to reach consensus about the final state of the block.
	
	In order to leverage the advances of multicores, this paper explores the next generation of smart contract platforms that enables concurrent execution of such contracts. Reasoning about the validity of the object states is challenging in concurrent smart contracts. We examine a programming model to support \emph{optimistic} execution of SCTs. We introduce a novel programming language, so-called \ov, and a Solidity API to ease programing of optimistic smart contracts. \ov\ language together with static checking will help reasoning about a crucial property of optimistically executed smart contracts -- the validity of object states in trustless systems.
\end{abstract}


\newpage
\pagenumbering{arabic} 
\tableofcontents 

\newpage

\section{Introduction}\label{ch:intro}

Engineering softwares in legacy computer systems is challenging, especially to maintain systems's properties in the existence of concurrent calls that mutate system's objects~\cite{philippsen00,lea:concurrentJava,goetz,matsuoka}. It is even more challenging to write smart contracts (SC) to be deployed on top of blockchains  due to the nature of trustless environment. 

As blockchain technologies and cryptographic algorithms have been more efficient and secure, smart contracts have become increasingly popular. There have been many types of smart contracts since the term was first introduced~\cite{szabo1997formalizing}.  Billions worth of USD are controlled by smart contracts world-wide, which are written in a number of programming languages to suite specific use cases of such programs. Blockchains rely on a consensus mechanism to reach agreement with peers on the sequential order of transactions in a block. Finalized transactions are committed as a group to the distributed ledger. All peers on the network perform a validation step by re-executing the transactions within the block and checking that the initial and final states match.

Ensuring correctness of smart contracts is a critical security concern as there have been exploitations of subtle bugs in these programs in the past couple of years. Another challenge is that SC code's immutability after deployment makes bug fixing impossible. Current audit practices of smart contracts involve checking whether the code is safe against two kinds of vulnerabilities: (i) generic security errors such as reentrancy and overflows, and (ii) deeper custom functional requirements; for example, "the sum of the deposits never exceeds the contract's balance". 
There have been a number of security tools for discovering generic errors and a few automatic verifiers of deeper functional requirements~\cite{permenev2020verx}.

In this paper, we consider smart contracts that are executed on blockchain-based platforms. We leverage the common concept of the object invariants (i.e., functional requirements of objects) to address the challenge of expressing and reasoning about object invariants in smart contracts.
Prior to describing our approach to allowing object validity of smart contracts to be specified and verified, we first highlight the importance of object validity in software development and then an overview of smart contracts.

\subsection{Validity and Program Reasoning}
A popular approach to enginering softwares is to model softwares with real-world objects or concepts that show clear program structures. Objects are often required to satisfy certain conditions --- object \emph{invariants} that must be maintained to guarantee the expected behaviour of the system. Any violation of the object invariant is often undesirable as it violates program logic and can cause programs to crash~\cite{jacobs2004vmo,jacobs2005,naumann2004tim}.
Here, this paper is not strictly limited to object-oriented programming, though we assume such O-O practices as much as possible.

Generally, reasoning about and maintaining object invariants in concurrent programs is difficult. The difficulties come from arbitrary object aliases, mutable object state, methods' side effects and the sophisticated mixture of multiple threads and shared mutable state typically of concurrent O-O programs.

\begin{lstlisting}[language=Java,multicols=2,frame=no,basicstyle=\small,	caption={An example of Customer and Account written in Java},label={fig:exposure}]
class Account { 
	private int amount = 0;
	// inv: amount >=0 
	
	int balance() { 
		return amount;
	}
	
	void deposit(int x) { 
		amount += x; 
	}
	
	void withdraw(int x) { 
		amount -= x; 
	}
}

class Customer {
	private Account a;
	private String id;
	// inv: a!= null && a.amount >= 10 
	
	void transfer(int amt, String custID) {
		Customer b = ....; // lookup custID
		a.withdraw(amt);
		b.deposit(amt);
	}	
	Account getAccount() { 
		return a;
	}
}
\end{lstlisting}

Listing \ref{fig:exposure} shows an example showing the challenges for maintaining object validity. A bank customer has an account. If a reference of account \textcode{a} of the customer is leaked, such as via a call \textcode{getAccount()}, then any manipulation of the reference is out-of-control of the customer. The alias  makes it very hard to reason about the invariant of the customer, which depends on the account. If such a manipulation, e.g. \textcode{withdraw}, on the leaked reference occurs concurrently with the client's operation, it would lead to a race condition on the account. As a result, no matter how carefully the Customer code is synchronized, neither a consistent state of the account nor the customer's invariant is guaranteed. Even worse, now suppose all the three operations of the account are manifested with lock-based synchronization (monitors or mutexes), to prevent the potential race condition. Unfortunately, \emph{deadlock} can occur if two customers attempt to simultanously transfer money to each other.

To mitigate the problems of locks, Software Transactional Memory (STM) systems \cite{Shavit1995stm} leverage a non-blocking approach promising to ease programming tasks for concurrent application developers. STM hides the complexity of concurrency control from programmers and enables safe composition of scalable applications.
STM offers a programming model inspired from the ACID properties (e.g. \emph{A}tomicity, \emph{C}onsistency, \emph{I}solation and \emph{D}urability) of database transactions which play a major role in software systems.
There are, however, few STMs ~\cite{Simon2006} paying attention to data invariants --- the \emph{consistency} guarantees.

For code deployed in trustless systems, it is even a harder problem to reason about validity of objects. Concurrent smart contracts are much more challenging to reason about, since execution is done via blockchain nodes in a trustless system.

\subsection{An Overview of Smart Contracts}
\emph{Smart contract} (SC), which was first proposed by Nick Szabo 20 years ago~\cite{szabo1997formalizing}, refers to computer code that automatically executes all or parts of an agreement and is stored on a blockchain-based platform. Smart contracts (SCs) are supplied as user-defined scripts to be executed on blockchain nodes. 
The SC code is replicated across multiple nodes of a blockchain and, therefore, benefits from the security, persistence and immutability that a blockchain offers. 
That replication also means that as each new block is added to the blockchain, the code is, in effect, executed. 
With recent blockchain technologies, smart contracts can handle more sophisticated transactions, with an increasing number of on-chain tokenized assets.
Developers may compose multiple transaction steps to form more complex smart contracts. 

Bitcoin Script is the first successful implementation of a blockchain-based smart contract, providing a purposely not-Turing-complete language with a set of simple, pre-defined commands. It allows a simple forms of SC for Bitcoin transactions, such as pay-to-public-key-hash (P2PKH) and pay-to-script-hash (P2SH). Recently, other platforms, such as Ethereum, can support more complex contractual functionalities and flexibilities. Solidity, which is a Turing-complete language for smart contracts developed by Ethereum, is one of the most popular languages for SCs.
Newer blockchain platforms, such as Hyperledger Fabric \cite{androulaki2018hyper}, allow smart contracts to be written in various high-level languages.

Conventional smart contract plaforms, such as Ethereum, leverage the \emph{sequential} execution model of transactions within a block. The platforms have enabled the execution of transactions of various complexity in blocks via user-defined smart contracts. Each block of the chain consists of multiple smart contract transactions (SCTs). These SCTs are executed \emph{sequentially} by the miner before appending the block into the miner's blockchain. The remaining peers, acting as validators, on receiving this block will again sequentially re-execute the SCTs of the block. The validators agree with the miner about the final state of the block on their records. After validated, the block is added to the blockchain.

There has been some research work studying concurrent execution of transactions in a block. 
For example, one such approach is based on object semantics by miner using STM~\cite{anjana2019efficient}. Concurrent miner stores a BlockGraph (BG) structure, which captures the object-conflicts relations among the SCTs, inside the block. Validators re-execute the same SCTs concurrently and deterministically, using the BG given by miner. The proposed block is appended into the blockchain if successful validation, or discarded otherwise. To address the double-spending issue, Optimistic Validator uses counter to find any malicious block proposed by a malicious miner~\cite{kokoris2018omniledger}. Using concurrent miners and validators can achieve better performance than existing SC frameworks.

\subsection{Our Approach}

Adding concurrency to the execution of SCTs can potentially help achieve better efficiency and higher throughput~\cite{anjana2019efficient,dickerson2019adding,zhang2018enabling}. Yet, there is no research in reasoning about object validity and security issues such reentrancy, in SCs on these platforms. In this paper, we are interested in the next generation of smart contracts platforms that can execute the SCTs concurrently to harness the power of multi-core processors. 
In order to address the security concern, we introduce a new approach that allows developers to specify validatity contracts in smart contracts. Valitidy contracts are used to statically checked against any violation of contract as well as reentrancy issues.

\subsubsection{Solidity API for Validity Contract Specifications}
There has been an extensive research in smart contracts tools and frameworks. But there is no research in how to specify object's validity in smart contracts. With Solidity, which is the most popular programming language for writing smart contracts, there is still a lack of a way to specify object's invarants~\cite{solidityInvariantReq}. 

To address this challenge, this paper introduces a new approach to allow programmers to specifying object's invarants into Solidity programs.
The new Validity API uses invariants for \emph{validity} guarantees of objects in SCs. 

\begin{minipage}{.45\textwidth}
	\begin{lstlisting}[language=sol,caption={An Storage contract in Solidity}, label={lst:storageSol},basicstyle=\small]
// file: storage.sol
pragma solidity >=0.5.16 <0.7.0;

/**
* @title Storage
* @dev Store & retrieve value in a variable
*/
contract Storage {
	uint256 number;
	
	/**
	* @dev Store value in variable
	* @param num value to store
	*/
	function store(uint256 num) public {
		number = num;
	}
	
	/**
	* @dev Return value 
	* @return value of 'number'
	*/
	function retrieve() 
			public view returns (uint256){
		return number;
	}
}
	\end{lstlisting}
	\begin{lstlisting}[language=ov,caption={An Storage contract in \ov}, label={lst:storageOV},basicstyle=\small]
// file: storage.ov		
class Storage [o] { 
	uint256 number;
	// inv: number > 0
	
	public void store(uint256 num) <this,this> {
		number = num;
	}
	
	public uint256 retrieve() <this,top> {
		return number;
	}
}
	\end{lstlisting}
\end{minipage}\hfill
\begin{minipage}{.5\textwidth}
	\begin{lstlisting}[language=sol,caption={An Storage contract converted from OV to Solidity. Storage's invariant is given in isValid(). Checking functions thisThis(), thisTop(), or botThis() are appended as function modifiers}, label={lst:storageOVSol},basicstyle=\small]	
// file: storage_ov.sol
pragma solidity >=0.5.16 <0.7.0;

import '../Ownable.sol';
import '../Validity.sol';

/**
* @title Storage
* @dev Store & retrieve value in a variable
*/
contract Storage_OV is Ownable, Validity {
	uint256 number;
	
	/**
	* @dev Store value in variable
	* @param num value to store
	*/
	function store(uint256 num) 
			preValid() postValid() public {
		number = num;
	}
	
	/**
	* @dev Return value 
	* @return value of 'number'
	*/
	function retrieve() preValid() 
			public view returns (uint256) {
		return number;
	}
	
	/**
	* @dev Return bool
	* @return true if the invariant holds; false otherwise
	*/ 
	function isValid() 
			external view returns (bool) {
		return number > 0;
	}
}
	\end{lstlisting}
\end{minipage}

In Listing~\ref{lst:storageOVSol}, methods are appended with a method \textcode{isValid()} that specifies the invariant of Storage contract. Method \textcode{store} is appended with new modifiers \textcode{preValid()} and \textcode{postValid()} to check pre-condition and post-condition. Function retrieve() has a precondition checked by appended modifier \textcode{preValid()}.

Although the discussions are in Solidity, it is worth-mentioning that the concept of our work is not restricted to Solidity; it can be extended to support other SC languages.

\subsubsection{Validity-Based Programming Language}

We introduce a novel language, called \ov~ (Optimistic valididy or Object validity), to help reason 
about the consistency in concurrent SC applications. Validity contracts for methods and transactions specify assumptions and effects on object invariants. The new language enables modular reasoning about object validity in the code and ensuring the validity of objects at runtime.

The contract-based programming model in the form of pre- and post- checks provides the system the consistency required for ACID transactions. All complex details of the underlying execution mechanisms (e.g., locks or transactions; the implementation details of STM library) are hidden from the programmers. With validity contract,  \ov\ system can reduce the number of required validity checks compared to what other approaches would require if they attempted to perform similar checks. 
Furthermore, validity contracts open a number of opportunities for optimization.

Listing~\ref{lst:storageOV} shows an example of Storage contract written in new OV language. Storage contract has an owner ("\textcode{[o]}") that is uniquely idenfitied at compile time and at runtime. Method \textcode{store} requires the Storage to be valid (first "\textcode{this}") and it can invalidate the Store's invariant inside its body (second "\textcode{this}"). Assertions of Storage's validity at the beginning and the end is required for the \textcode{store} method. Method \textcode{retrieve} requires the Storage's invariant, but does not modify the state of Storage and so it invalidates nothing (pseudo "\textcode{top}"). By using the validity contracts ("\textcode{<this,this>}" and "\textcode{<this,top>}"), static checks can get rid of illegal operations.

Listing~\ref{lst:ballot} gives a code fragment of a Ballot contract writing in Solidity. Listing~\ref{lst:ballotOV} expresses the concept of Ballot contract's owner ("[o]") and validity contract ("\textcode{<this,this>}"). The validity contract specifies that method vote requires the Ballot contract is valid (and so all of its owned contracts are valid). The validity contract also states that the method vote may make some modifications to the ballot's state, 

\noindent\begin{minipage}{.47\textwidth}
	\begin{lstlisting}[language=sol,caption={A Ballot contract in Solidity}, label={lst:ballot}]
contract Ballot {
	mapping(address => Voter) public voters;
	// more state definitions
	function vote(uint proposal) {
		Voter sender = voters[msg.sender];
		if (sender.voted)
			throw;
		sender.voted = true;
		sender.vote = proposal;
		proposals[proposal].voteCount += sender.weight;
	}
	// more operation definitions
}
	\end{lstlisting}
\end{minipage}\hfill
\begin{minipage}{.47\textwidth}
	\begin{lstlisting}[language=ov,caption={A Ballot contract in \ov}, label={lst:ballotOV}]
class Ballot[o] {
	public map<this> voters; // address => Voter
	// more state definitions
	void vote(uint proposal)<this,this> {
		Voter sender = voters.get(msg.sender);
		if (sender.voted)
			throw;
		sender.voted = true;
		sender.vote = proposal;
		proposals[proposal].voteCount += sender.weight;
	}
	// more operation definitions
}
	\end{lstlisting}
\end{minipage}

The benefits of validity contracts are three-fold. First, validity contracts are served for detecting transactional conflicts at a higher abstraction level (e.g., \emph{validity interference}) than reads and writes. In STM, \emph{conflict detection} determines when two transactions cannot both safely commit, to guarantee serializability.
Second, validity contracts can be used for conflict resolution in \emph{contention management} to guarantee progress (livelock avoidance)~\cite{herlihy2003dstm,SXM2005}, to increase throughput dramatically in high contention workloads ~\cite{SCOOL2005,PODC2005}.
With validity contracts, contract-based contention managers with different safety-driven strategies can be developed to ensure liveness.
Third, validity checks can be deduced from contracts and scheduled at transaction commits to preserve object validity. Reentrancy issues can be avoided from enforcing the validity requirements and effects of operations inside a transaction's body. Validity contracts can help reduce the number of validity checks while ensuring safety. The validity contracts can be used to insert correct assertions of validity, e.g., calling \textcode{preValid()}, \textcode{postValid()} from the Validity API.
\emph{Transactional validation} ensures a transaction never views or makes use of inconsistent data; while \emph{object revalidation} ensures transactions only commit valid data.

\subsection{Contributions}\label{sec:contri}

This paper makes the following contributions:
\begin{enumerate} 
	\item [$-$] {\bf Validity API} We present a new API that provides a feasible solution to allow developers to specify object validity in smart contracts written in Solidity.
	\item [$-$] {\bf New Programming Model for  Validity Contracts:}
	This paper proposes a general programming model to allow developers to specify validity contracts for smart contracts. The validity contracts ensure sufficient checks for validity requirement at a function call start. The validity contracts are useful for detecting and scheduling conflicting transactions, and for software assurance by automating object revalidation via deduced assertions.
	\item [$-$]{\bf Reasoning about Optimistic Smart Contracts:} 
	This work is the first that introduces a practical way to express validity contracts in smart contracts. We present a new language \ov~ integrating ownership mechanism, type system and dynamic semantics for modular reasoning about object validity and validity contracts for smart contract transactions. The validity contracts are useful to enhance program safety and to improve modularity and composability of optimistic smart contracts.
\end{enumerate}

\subsection{Paper structure}

The rest of this paper is organised as follows. 
Section~\ref{se:api} introduces our new Validity API written in Solidity and an example of validity contracts using our API. 
Section~\ref{se:model} presents our validity-aware model and new \ov\ language for optimistic smart contracts.
The section gives the execution model, abstract syntax and semantics of the new language.
Section~\ref{se:discuss} gives some discussions about possible optimization using validatity contracts.
Section~\ref{se:related} gives an overview of the related work.
Section~\ref{se:con} concludes.

\section{New Solidity API for Validity Contracts}\label{se:api}

To overcome the shortcomings of Solidity, we introduce a new 
Validity API to allow developers to specify and reason about object validity.

\subsection{Validity API}

Ownable contract describes a general kind of contracts that has an owner. There is a private field '\textcode{owner}' that will be initiated at contract creation; caller that creates a contract is the owner of the created contract.
Modifier \textcode{isOwner} is used in functions whose invocation is restrictive to the contract's owner.
Modifier \textcode{isCalledBy} is dedicated for functions that are called from a particular caller's address.

\begin{lstlisting}[language=sol,caption={An Ownable contract in Solidity}, label={lst:ownableSol},basicstyle=\small]
// file: ownable.sol
pragma solidity >=0.5.16 <0.7.0;

/**
* @title Ownable
* @dev Set and get owner
*/
contract Ownable 	
	// modifier to check if caller is owner
	modifier isOwner() {
		require(msg.sender == owner, "Caller is not owner");
		_;
	}
	
	// modifier to check if caller is owner
	modifier isCalledBy(address addr) {
		require(msg.sender == addr, "Caller is not the specified address");
		_;
	}
		
	// @dev Set contract deployer as owner
	constructor() public {
		owner = msg.sender; // 'msg.sender' is sender of current call
	}
	
	/**
	* @dev Return owner address 
	* @return address of owner
	*/
	function getOwner() external view returns (address) {
		return owner;
	}
	
	address private owner;
}
\end{lstlisting}

In this API, ownership transfer is not allowed. Each contract has a unique immutable owner at creation time. This is to guarantee the ownership tree (which-owns-which) is unique to ease the reasoning of validity of SCs.

Validity interface gives a simple interface that defines \textcode{isValid()} function and two modifiers. Actual contracts that implement Validity interface must specify its own implementation of \textcode{isValid()} function. Modifier \textcode{preValid()} is used for a function that require the contract's validity before the function's invocation. Whereas modifier \textcode{preValid()} asserts a post-condition of  the contract's validity after the function is called.

\begin{lstlisting}[language=sol,caption={An Storage contract in \ov}, label={lst:validaitySol},basicstyle=\small]
// file: validity.ov		
pragma solidity >=0.5.16 <0.7.0;

/**
* @title Validity
* @dev define validity of an object
*/
interface Validity {	
	/**
	* The invariant condition of an object.
	* Subclass must implement this method speciyfing its invariant.
	*/ 
	function isValid() external view returns (bool);
	
	// modifier to check object's validity prior a function call
	modifier preValid() {
		require(this.isValid(), "Validity fails pre-check");
		_;
	}
	
	// modifier to check object's validity immediately after a function call
	modifier postValid() {
		_;
		require(this.isValid(), "Validity fails post-check");
	}
}
\end{lstlisting}

OVValidity interface extends Validity interface by adding several extra modifiers. These additional modifiers are used to ease the transformation of OV program into Solidity.
Modifier \textcode{thisThis()} is used for functions that require pre- and post- checks of a contract's validity. This corresponds to the contract \textcode{<this,this>} in OV.
Modifier \textcode{botThis()} is used for functions that require a post- check of a contract's validity. This corresponds to the contract \textcode{<bot,this>} in OV.
Modifier thisTop() is used for functions that require a pre- check of a contract's validity. This corresponds to the contract \textcode{<this,top>} in OV.
Modifier \textcode{botTop()} is used for functions that neither require a pre- or post- check of a contract's validity. This corresponds to the contract \textcode{<bot,top>} in OV. This modifier can be safely ignored, as functions with this validity contract does not require such checks.

\begin{lstlisting}[language=sol,caption={An OVValidity interface specifying validity function isValid() and modifiers thisThis(), thisTop(), or botThis() are appended as function modifiers}, label={lst:ovvaliditySol},basicstyle=\small]	
// file: ovvalidity.sol
pragma solidity >=0.5.16 <0.7.0;

/**
* @title OVValidity
* @dev define validity of an object
*/
interface OVValidity {
	/**
	* The invariant condition of an object.
	* Subclass must implement this method speciyfing its invariant.
	*/ 
	function isValid() external view returns (bool);
	
	// modifier to check object's validity prior a function call
	modifier preValid() {
		require(this.isValid(), "Validity fails pre-check");
		_;
	}
	
	// modifier to  object's validity immediately after a function call
	modifier postValid() {
		_;
		require(this.isValid(), "Validity fails post-check");
	}
		
	// The following modifiers are short-hand for OV language
	
	// modifier to check object's validity before and after a function call
	modifier thisThis() {
		require(this.isValid(), "Validity fails pre-check");
		_;
		require(this.isValid(), "Validity fails post-check");
	}
	
	// modifier to check object's validity after a function call
	modifier botThis() {
		_;
		require(this.isValid(), "Validity fails post-check");
	} 
	
	// modifier to check object's validity before a function call
	modifier thisTop() {
		require(this.isValid(), "Validity fails pre-check");
		_;
	}
	
	// modifier that is simply not checking object's validity at all
	modifier botTop() {
		_;
	} 
}
\end{lstlisting}

\subsection{Example}
We then present an example of smart contracts using our Validity API. Listing~\ref{lst:accountSol} shows an Account contract writing in Solidity. An Account contract has a balance and three functions. The account balance has an invariant that requires the balance is greater than 0 and less than $1e30$. Functions deposit and withdraw can modify the balance field and thus can invalidate the account's invariant. Function get simply retrieves the value of the balance.

Listing~\ref{lst:accountOV} shows the \textcode{Account} contract written in OV language. The account has an owner ("\textcode{[o]}").
Functions \textcode{deposit()} and \textcode{withdraw()} require the account's invariant to be valid before and after the function's body. In OV, the validity contracts are specified as \textcode{<this,this>}.
Function \textcode{get()} retrieves the balance without any modification to the account's state (the balance field). Validity contract specified in OV as \textcode{<this,top>}.

Listing~\ref{lst:accountOVSol} gives a program, which is the result of a transformation from OV to Solidity. The account contract implements \textcode{Ownerable} contract and \textcode{OVValidity} interface. The object validity is defined in isValid function. The \textcode{isValid} function itself does not require any validity checks and thus \textcode{botThis()} modifier is used.
Functions \textcode{deposit()} and \textcode{withdraw()} are appended with modifier \textcode{thisThis()}. Function get is appended with modifier \textcode{thisTop()}.

\begin{minipage}{.47\textwidth}
	\begin{lstlisting}[language=sol,caption={An Account contract in Solidity}, label={lst:accountSol},basicstyle=\small]
pragma solidity >=0.5.16 <0.7.0;
/**
* @title Account
* @dev Deposit, withdraw and get balance
*/
contract Account {
	uint256 balance;	
	// invariant: balance > 0 && balance < 1e30
	
	/**
	* @dev Deposit an amount to the account
	* @param amount value to deposit
	*/
	function deposit(uint256 amount) public {
		balance += amount;
	}
	
	/**
	* @dev Withdraw an amount from the account
	* @param amount value to withdraw
	*/ 
	function withdraw(uint256 amount) public {
		balance -= amount;
	}
	
	/**
	* @dev Return value 
	* @return value of the account
	*/
	function get() public view returns (uint256){
		return balance;
	}
}
	\end{lstlisting}
	\begin{lstlisting}[language=ov,caption={An Account contract in \ov}, label={lst:accountOV},basicstyle=\small]	
class Account [o] { 
	uint256 balance;
	// inv: balance > 0 && balance < 1e30
	
	public void deposit(uint256 balance)<this,this> {
		balance += amount;
	}
	
	public void withdraw(uint256 balance)<this,this> {
		balance -= amount;
	}
	
	public uint256 get()<this,top> {
		return balance;
	}
}
	\end{lstlisting}
\end{minipage}\hfill
\begin{minipage}{.51\textwidth}
	\begin{lstlisting}[language=sol,caption={An Account contract converted from OV to Solidity. Account's invariant is given in isValid(). Checking functions thisThis(), thisTop(), or botThis() are appended as function modifiers}, label={lst:accountOVSol},basicstyle=\small]	
pragma solidity >=0.5.16 <0.7.0;

import '../Ownable.sol';
import '../OVValidity.sol';

/**
* @title Account
* @dev Deposit, withdraw and get balance
*/
contract Account is Ownable, OVValidity {
	uint256 balance;	
	// invariant: balance > 0 && balance < 1e30
	
	/**
	* @dev Deposit an amount to the account
	* @param amount value to deposit
	*/
	function deposit(uint256 amount) thisThis() public {
		balance += amount;
	}
	
	/**
	* @dev Withdraw an amount from the account
	* @param amount value to withdraw
	*/ 
	function withdraw(uint256 amount) thisThis() public {
		balance -= amount;
	}
	
	/**
	* @dev Return value 
	* @return value of the account
	*/
	function get() thisTop() public view returns (uint256){
		return balance;
	}
	
	/**
	* The invariant 
	* @dev Return bool
	* @return true if the invariant holds; false otherwise
	*/ 
	function isValid() botThis() 
			external view returns (bool) {
		return balance > 0 && balance < 1e30;
	}
}
	\end{lstlisting}
\end{minipage}

\section{\ov: Validity-Aware Optimistic Smart Contracts}\label{se:model}

We present our new programming model, in which transaction has a validity contract specifying the requirements and effects.  
Transaction contracts are useful for program verification and optimization.
Based on the design-by-contract (DbC) discipline, transaction contracts can be used to ensure system validity by supporting both static or dynamic checking, and can be used to help detect and resolve conflicts between transactions.

\subsection{Validity Contracts for Smart Contract Transactions}
 \label{sec:contract}

Transaction comprises a set of operations to be executed all at once, or none at all. Transactions often preserve the ACI (atomicity, consistency, isolation) properties. A transaction may perform operations on shared objects, as well as local data (objects) that are inaccessible to other transactions.
Each transaction, distinguished by a unique identifier, is initially live and may eventually become either committed (successful) or aborted (failed or cancelled).

Our new language, namely \ov, generalizes the idea of validity contracts and validity subcontracting that are introduced in AVID~\cite{nguyen2009avid} for contract specifications of STM transactions in legacy softwares. Specifying the validity contracts in \ov\ is similar to read-write effects in previous work \cite{lu07ecoop,Clarke98Ownership,clarke01simple,david02ownership} for sequential programs.
Here, \ov\ leverages the idea of \emph{validity contract} to describe behavioural abstraction of a SC transaction.

 Listing~\ref{fig:contract} describes the underlying mechanism of validity contracts for SC transactions. A validity contract of a SC transaction, written as $\modelVI$, specifies the validity requirements and effects of the transaction. The \emph{validity set} $\modelV$ specifies a set of objects that must be valid both before and after transaction execution.
The \emph{invalidity set} $\modelI$ specifies a set of objects that may be invalidated by the execution of the transaction.
These objects in $\modelI$ are the only ones whose fields may be modified within the transaction's execution.
That means the invariant of the objects within $\modelV \cap \modelI$ may be broken during the transaction execution.
These objects must be revalidated prior to transaction commit; all objects in the validity set are ensured valid once transaction commits.

\begin{center}
\begin{minipage}{.9\textwidth}
	\begin{lstlisting}[mathescape=true,
	caption={Meaning of Transaction's Validity Contract},label={fig:contract}]
Validity contract $\modelVI$ of a transaction $\tran{T}$ implies:
    Begin: requires $\modelV$ valid
    Executing: may modify $\modelI$ 
    Prior to commit: revalidate $\modelV\cap\modelI$
    Finish: ensures $\modelV$ valid; $\modelI\setminus\modelV$ may still be invalid
	\end{lstlisting}
\end{minipage}
\end{center}

\emph{Validity subcontracting} restricts which operations may be performed by a specific contract. Subcontracting constrains the invariant requirements and effects of calls or subtransactions nested inside a transaction or a method, such as the relationship between a caller and each of its callees, and between a transaction and each of its nested transactions. That is, $\modelVIX$ is a subcontract of $\modelVI$ if $\modelV' \subseteq \modelV\ {\rm and}\ \modelI' \subseteq  \modelI$. Like AVID, the version of subcontract in \ov\ allows more programs to be type checked, but requires additional runtime validity checks. 

A context $V$ in \ov~ is a syntactic sugar for the set of objects in the subtree of the ownership tree rooted at $V$ (as in \cite{lu07ecoop,david02ownership}). Ownership contexts are used to specify \emph{both} validity and invalidity sets. A \textcode{bot} context represents an empty set. A transaction/call with \textcode{bot} validity set implies no validity check required at start. A \textcode{bot} invalidity set implies the transaction/call neither modifies nor invalidates any object.

\ov\ transactional execution mechanism allows roll back of a transaction that causes violation of some object's invariant.
Pure queries having an empty $\modelI$ and so require no revalidation.
The overlap of $\modelV'$ and $\modelV\setminus\modelI$ specifies the set of objects whose validity is required at the start of the (sub)transaction/call. No revalidation is necessary at the end of a transaction when there is no overlap between $\modelV$ and $\modelI$.

\subsection{Execution Model}
Execution model of \ov\ programs allows a transaction to create a sub-transaction by using \textcode{atomic\{S\}} expression. Semantics for contract-attached atomic expression in our model is similar to the one used in \cite{herlihy2003dstm,fortress,chapel,x10}. 
\ov\ transaction may nest a number of sub-transactions.
Here, we assume a simpler model, called \emph{linear nesting} \cite{Moss2006,IBM2008}, which restricts a transaction to have at most one live child. This simple form ensures atomicity of the nested transactions and validity of the objects rather than full concurrency optimization. Top-level transactions can run concurrently and commit their effects to the memory.
Generally speaking, \ov's general execution model can support \emph{closed} nested transactions \cite{Moss1981}, in which all memory locations being accessed by nested transactions are conceptually considered as being accessed by its parent transaction. When a closed-nested transaction commits, its read and write sets are merged with those of its parent.

\emph{Validity requirement} specifies how transactions can guarantee the validity at transaction start. A transaction may wait for validity prior to transaction start and may revalidate validity of an object before it commits. For example, by using conditional atomic regions \cite{carlstrom2006atomos, chapel,x10} or conditional critical regions (CCR)~\cite{Harris2003}. The \textcode{when(o.valid())\{S\}} expression waits until the object invariant being held before it starts the transaction execution. 

\emph{Transaction validation} ensures the consistency of the objects accessed by a transaction, with respect to other concurrent transactions. 
Retrying an aborted transaction is considered as executing a new transaction with same validity contracts, although with a different transaction identifier. 
\emph{Object revalidation} ensures the consistency of the objects according to the contract of the particular transaction. 
In \ov's general model, transaction performs an automatic check for the validity of the object in the transaction validation step. The validity check is deduciable from the transaction's contract.

\emph{Validity subcontract} generalises a similar concept in AVID~\cite{nguyen2009avid} used only for concurrent programs, to apply for smart contract transactions.
For all method definitions, it guarantees all expressions inside a method are covered by the method contract.
For all transactions, subcontracting is applied to parent and child transaction, or to a transaction and its nested calls/transactions. 
Parent transaction must ensure the validity set of a subtransaction's contract before it starts. The invalidity set of the subtransaction must be captured within that of its parent transaction.

\subsection{Language} \label{sec:lang}

Here, we present our \ov~language, which is a Java-like language, for secure smart contracts. \ov~language is based on \avid~language ~\cite{nguyen2009avid}, which was introduced for STM in legacy softwares. More details about the validity contracts, type systems and semantics can be found in AVID~\cite{nguyen2009avid}.

The abstract syntax of \ov\ language is shown in Table \ref{table:source}. 
The type system relies on ownership types for managing dependencies between object invariants. 
The metavariables $T$ ranges over types; $K$ ranges over contexts; $V,I$ ranges over validity and invalidity sets ; $P$ ranges over programs; $L$ ranges over class definitions; $M$ ranges over method definitions; $e$ ranges over expressions; $C$, $X$, $m$, $f$ and $x$ range over class, formal context parameters, method and field names, and variables names, respectively. \keyword{this} refers to the target object for the current call and is also used as an ownership context.
The overbar is used for a possibly empty sequence of constructs; for example, $\overline{e}$ is used for a finite ordered sequence $e_{1} .. e_{n}$, $\overline{T}\ \overline{x}$ stands for a possibly empty sequence of pairs $T_{1}\ x_{1} .. T_{n}\ x_{n}$.
An expression $e$ can be either a constant $c$, a variable $x$, the $\keyword{this}$ pseudo variable, a field access $e.f$,
a method invocation $e.m(\overline{e})$, an object construction $new\ T$, a sequential composition ($\keyword{seq}(e,e)$), a thread creation ($\keyword{fork}\ e$) and an atomic block ($\keyword{atomic}[\Delta]_{opt}\ e$).

\begin{table}[htb]\small\centering
	\begin{tabular}[c]{| lrl  @{\hspace{56pt}} l|}
		\hline
		$T$&$::=$&$C\langle \overline{K}\rangle\ |\ \keyword{int}\ |\ \keyword{bool}$&types\\
		$K$&$::=$&$X\ |\ \keyword{top}\ |\ \keyword{bot}\ |\ \keyword{this} \ |\ *$&contexts\\
		$V,I$&$::=$&$K$&validity and invalidity sets\\
		$P$&$::=$&$\overline{L}\ e$&programs\\
		$L$&$::=$&$\keyword{class}\ C[\overline{X}]\lhd T\ \keyword{where}\ \overline{X R X}$ & \\  
		&& $\{\overline{[\keyword{final}]_{opt}\ T\ f};\ \overline{M}\}$& classes\\
		$R$&$::=$&$\prec\ |\ \preceq $&constraints\\
		$M$&$::=$&$\methsyntax$&methods\\
		$\Delta$&$::=$&$\langle V,I\rangle$&contracts\\
		$e$&$::=$& $c\ |\ v\ |\ \keyword{new}\ T\ |\ x=e$ & \\
		&& $|\ v.f \ |\ v.f= v \ |\ v.m(\overline{x})$ & \\
		&& $|\ \keyword{seq}(e,e) \ |\ \atomic\ [\Delta]_{opt}\ e\ |\ \keyword{fork}\ e $ & expressions\\
		$v$&$::=$&$\keyword{this}\ |\ x$&values\\
		$C$&$::=$&$id$&class names\\
		$X$&$::=$&$id$&formal ownership parameter names\\
		$m$&$::=$&$id$&method names\\
		$f$&$::=$&$id$&field names\\
		$x$&$::=$&$id$&variable names\\
		\hline
	\end{tabular}
	\caption{Abstract Syntax} \label{table:source}
\end{table}

In \ov\ language, threads can only be created at top-level, not within a transaction, as in existing proposals including \cite{moore2008popl}. 
The main constructs of the language are two constructs: \emph{atomic} for transaction and \emph{fork} for concurrency. Effect clauses are used to name object sets as required for validity contracts. An \ov\ atomic block has a validity contract that is optionally specified, i.e.,  ($\keyword{atomic}[\Delta]_{opt}\ e$) for executing $e$ as a transaction. The optional contract $\Delta$ is supplied to an $\atomic$ expression in case the contract cannot be deduced, i.e., for a code block; whereas no explicit contract is required in the case of a single expression $e$ (such as a single assignment or method call).
A transaction, with contract \valcontract{this,this}, must revalidate \keyword{this} object prior to commit.
\ov~language does not provide explicit abort/retry construct, although this construct are easily added as in prior work \cite{harris2005exception} with no evaluation rule for it.

Programs consist of a collection of classes with an expression $e$ to be evaluated.
Classes are parameterised with context parameters having some constraints expressed by the \keyword{where} clause such as assumed ordering.
The first formal context parameter of a class determines the owner of \keyword{this} object within the class. Contexts can be formal class parameters; the current context \keyword{this}; the root \keyword{top} and the bottom \keyword{bot} of the ownership hierarchy.
The actual owner of an object is determined by its creation type, and is fixed for the lifetime of the object.
For the sake of simplicity, the expression of object invariants and their dependence on fields is not formalized. Instead, object's invariant must only depend on its own fields and on fields of other objects that it (transitively) owns.

As an example, Listing \ref{lst:accountOv} shows the bank account contract example, which is written in \ov\ abstract syntax.
Ownership types and validity contracts are highlighted in the example.

	\begin{lstlisting}[language=ov,
	caption={Account and Customer example in OV abstract syntax},label={lst:accountOv},multicols=2,basicstyle=\small,frame=no]
class Account[o] {
	int amount = 0;
	// inv: amount >= 0	
	Account(int amt) { 
		amount = amt;
	}	
	int balance() <this,bot> { 
		return amount; 
	}	
	void deposit(int x)<this,this>{ 	
		amount += x;	
	}	
	void withdraw(int x)<this,this>{ 
		amount -= x; 
	}
}
	
class Customer {
	Account<this> a = new Account<this>(0);
	// inv: a != null && a.amount >= 10 
	
	void safeWithdraw(int amt)<this,this>{
		verifyLogin(); 
		...
		atomic a.withdraw(amt);
	}
	
	void verifyLogin()<bot,this>{ 
		... 
	}
}
	\end{lstlisting}

Transactions are created explicitly with \keyword{atomic} keyword. \ov\ transactional  framework  will take as input the Java-annotated programs and schedule the transactions.
The specified validity contract \valcontract{this,this}  in the Account's \textcode{withdraw} method has two usages. First, it helps detect possible interference with transactions. Second, my system can deduce that account \textcode{a} needs to be revalidated prior to committing the transaction. My transactional framework will perform 
the validity check by invoking \textcode{a.valid}. The transaction will commit if the check returns true; otherwise, the transaction will be aborted and retried. This safety guarantee is the ultimate goal of my work.

\subsection{Type System}\label{sec:types}

This section gives most relevant rules of the \ov\ language, including contract checking. The full type system is included in ~\cite{nguyen2009avid}. 
Table~\ref{table:syntax2} extends the abstract syntax with type environments for use in the type system. I define an existential context ? to be included in context $K$. 
I define a typing environment $\Gamma$ as $\bullet\ |\ \Gamma,X\ |\ \Gamma,X\ R\ X\ |\ \Gamma,v:T$. The typing environment records context parameters, the assumed constraints between context parameters, and the types of variables. Expression $\valid\ e$ is a validity assertion.

\begin{table}[h]\centering
	\begin{tabular}[c]{| lrl @{\hspace{38pt}}l |}
		\hline
		$K$&$::=$&$...\ |\ ?\ $&contexts\\
		$e$&$::=$&$...\ |\ \valid\ e $&expressions\\
		$\Gamma$&$::=$&$\bullet\ |\ \Gamma,X\ |\ \Gamma,X\ R\ X\ |\ \Gamma,v:T$&environments \\
		\hline
	\end{tabular} \caption{Extended Syntax for Type System} \label{table:syntax2}
\end{table}

Table \ref{table:contract} defines well-formed contract and subcontract rules for enforcing
validity subcontracting. The \rulename{contract-sub} rule specifies containment of contract requires both containment of validity requirements and invalidity effects. It also highlights the fact that calls that affect the validity invariant can be only originated from the owner context. The \rulename{effect-sub} states that sub-context implies sub-effect. Sub-effect is a transitive relation, specified by \rulename{effect-tra} rule.
The table also covers containment of context sets.

\begin{table}[h]\small
	\setlength{\tabcolsep}{1pt}
	\begin{tabular}{lll}
		\ruledef{Contract}{\context{V} \qquad \context{I}}{
			\contract{\langle V, I\rangle}}%
		\qquad 
		\ruledef{Contract-sub}{
			\subsetcon{V'}{V} \qquad \subsetcon{I'}{I}
		}{\subcontract{\langle V,I \rangle}{\langle V',I' \rangle}}%
		\qquad
		\ruledef{effect-sub}{ \insideeq{K}{K'} }{
			\subsetcon{K}{K'}}%
		\qquad 
		\ruledef{effect-tra}{ 
			\subsetcon{I}{I'} \qquad
			\subsetcon{I'}{I''}
		}{\subsetcon{I}{I''}}%
	\end{tabular} \caption{Rules for Validity Contract and Subcontract} \label{table:contract}
\end{table}

Table~\ref{table:type} provides rules for type well-formedness, subtyping rules for expressible types and
the rules for context abstraction. A separate judgement is introduced for bindability to handle types with existential contexts (which only occur in the type system after field or method lookup). For expressible types, there is no difference between subtyping and bindability. 
\rulename{abs-rfl} ensures that the existential contexts abstract nothing by insist they must be valid contexts.
Combined with \rulename{bin-abs} this ensures that
bindability is not reflexive, because existential types can only occur on the right-hand side of the binding relation. Through the use of bindability in typing field assignment and method argument binding, this prevents existential contexts from being associated with the target of a binding.

\begin{table}[h] \small
	\setlength{\tabcolsep}{1pt}
	\begin{tabular}{llll}
		\ruledef{typ-obj}{
			defin(C\langle\overline{K}\rangle,\_)=\keyword{class}\ C\ ...\
			\keyword{where}\ \overline{K'\ R\ K''}\\
			\context{\overline{K}} \qquad \constraintM{K'}{K''}
		}{\typeO{C\langle\overline{K}\rangle}}%
		\quad 
		\ruledef{type}{
			\typeO{T'} \\ \subtype{T'}{T}
		}{\type{T}}%
		\quad 
		\ruledef{sub-ext}{
			\keyword{class}\ C[\overline{X}]\lhd T'\ ... \\
			T=[\overline{K/X}]T'
		}{\subtype{C\langle\overline{K}\rangle}{T}}%
		\quad 
		\ruledef{sub-rfl}{ \\
		}{\subtype{T}{T}}%
		\qquad
		\\[30pt]
		\ruledef{sub-tra}{\subtype{T}{T'} \\ \subtype{T'}{T''}
		}{\subtype{T}{T''}}%
		\quad 
		\ruledef{bin-abs}{ \abstractionM{K}{K'}
		}{\binding{C\langle\overline{K}\rangle}{C\langle\overline{K'}\rangle}}%
		\qquad 
		\ruledef{bin-sub}{ \subtype{T}{T''} \\ \binding{T''}{T'}
		}{\binding{T}{T'}}%
		\qquad 
		\ruledef{abs-any}{\\
		}{ \abstraction{K}{*} }%
		\qquad 
		\ruledef{abs-rfl}{\context{K} 
		}{ \abstraction{K}{K} }%
	\end{tabular} \caption{Rules for Type, Subtype, Binding and Abstraction} \label{table:type}
\end{table}

\subsection{Dynamic Semantics and Properties}\label{sec:dynamic}

The operational semantics of \ov~ type system based on \avid~\cite{nguyen2009avid} are briefly presented here. \ov's transactional memory semantics is based on the semantics of \avid, which is built on top of Oval~\cite{lu07ecoop} and \emph{StrongBasic} language~\cite{moore2008popl}. 
Object validity can be ensured from validity contracts and the strong isolation of transactions.

Table \ref{table:dynsyntax} gives the extended syntax and features used by \ov\ dynamic semantics.
A program state has the form of $\alpha;H;\Sigma;S;\Pi$ where $\alpha$ indicates whether any thread is currently executing a transaction ($a=\circ$ for no and $\alpha=\star$ for yes), $H$ is a heap, $\Sigma$ is a set of valid objects in $H$, $S$ is a stack frames of contracts, and $\Pi$ is a collection of threads.
A heap $H$ is a mapping from locations $l$ to objects $o$; an object maps its fields $f$ to locations $l$.
All locations are annotated with the type of the object they refer to; but we may omit them wherever that information is not used. Values are extended with locations, thereby allowing locations to be used as ownership contexts.
At runtime, new expression $\keyword{inatomic}(e)$ is used to present a partially completed transaction with remaining computation $e$.
Let $\Pi_1 \|\ \Pi_2$ denote a combination of two thread collections into a larger collection, where operator $\|$ is commutative and associative.

\begin{table}[h] \small\centering
	\setlength{\tabcolsep}{1pt}
	\begin{tabular}[c]{lrl@{\hspace{15pt}} l}
		$l,l_T$&&&typed locations\\
		$v$&$::=$&$...\ |\ l$&values\\
		$e$&$::=$&$...\ |\ \keyword{inatomic}(e)$&expressions\\
		$o$&$::=$&$\overline{f\mapsto l}$&objects\\
		$H$&$::=$&$\overline{l\mapsto o}$&heaps\\
		$\Sigma$&$::=$&$\keyword{bot}, \overline{l}$& valid objects\\
		$\Pi$&$::=$&$\bullet\ |\ \Pi\ \|\ e$& thread collections\\
		$S$&$::=$&$\overline{\Delta}$& stack frames \\
		$\alpha$&$::=$&$\circ \ |\ \star $& thread existence conditions
	\end{tabular}
	\caption{Extended Syntax for Dynamic Semantics} \label{table:dynsyntax}
\end{table}

Like \avid, \ov's semantics is based on \emph{StrongBasic} language \cite{moore2008popl} and allows at most one thread to execute a transaction at a time, as if all transactions were protected by a single global lock \cite{Menon2008globallockSTM}. In this semantics, $\alpha$ acts like a global lock where $\star$ indicates the lock is held, to enforce atomicity and isolation.
Let $\bullet$ denote an empty entity, such as the initial heap, thread collection and the type environment which is not available in the dynamic semantics. Unused $\bullet$ can be omited, e.g., $\exprE{e}{T}$ in place of $\exprEX{\bullet}{e}{T}$; and $e$ in place of $\bullet\ \|\ e$. 
The source program $\overline{L}\ e$ contains only a single $e$, but the evaluation of $e$ can spawn new threads, which in turn can spawn new threads. 

Ownership information and validity contracts are needed in static type checking, and may be erased after type checking.
They are only used in a specific implementation such as for conflict resolution. Thus, they do not affect how expressions are evaluated; generally they can be erased once validity checks ($\valid\ e$) are inserted into translated program. However, to keep a connection between static and dynamic semantics, the dynamic semantics still include ownership and contracts.

We now present some of the key properties of the \ov's type system.
In this formalization, it is presumed that every transaction with a contract \valcontract{l,l} ending with $\valid\ l$ to allow abort/retry (i.e., undo any write $l'.f=...\ $ where $\insideeq{l'}{l}$) in the case when the check fails (*).
Important lemmas are given as follows. The proofs of these lemmas are described in \avid~\cite{nguyen2009avid}.

\begin{lem}[Progress] \label{lemma:progress} 
	Given \ $\Gamma\vdash \alpha;H;\Sigma;S;\Pi$, then either $\Pi$ is all values or $\exists\alpha';H';\Sigma';S';\Pi'$ such that $\reduce{\Pi}{\alpha'}{H'}{\Sigma'}{S'}{\Pi'}$.
\end{lem}
\textbf{Proof.}
If $\Pi$ is all values, then no reduction rule applies. Otherwise, the proof proceeds by induction on the form of
$\reduce{e}{\alpha'}{H'}{\Sigma'}{S'}{e';\Pi}$.

\begin{lem}[Type and Contract Preservation] \label{lemma:preservation} \ \\
	If\ \
	$\left\{\begin{array}{l}
	\validity\ \quad and\ \heap  \\
	\exprX{\bullet}{e}{T}
	\end{array} \right.$\ and\ $\reduceX{\alpha;H;\Sigma;S;e}{\alpha';H';\Sigma';S';e';\Pi}$, 
	then\ \ $\left\{\begin{array}{l}
	\validityX{H'}{\Sigma'}{\Delta'} \quad and\ \ \heapX{H'} \\
	\exprXX{\bullet}{\Delta'}{e'}{T}\quad and\ \binding{T'}{T}
	\end{array} \right.$\\ \\
	where $S=..., \Delta$ and $S'= ..., \Delta'$.
\end{lem}

\begin{lem}[System Validity Preservation] \label{lemma:validity}
	Given\ $\vdash \overline{L}\ e$\ and\ $\reduceX{\circ;\bullet;\bullet;$\valcontract{top,top}$;e}{\circ;H;\Sigma;S; v_1\ \|\ .. \ \|\ v_n}$,
	then $\Sigma = dom(H)$.
\end{lem}

\section{Discussions}\label{se:discuss}

This section gives several discussions about our approach to object validity in optimistic smart contracts.
Besides static checking, validity contracts are useful for optimistic execution optimization.

\paragraph{Validity Interference Dection} Validity contracts can be used for detecting transactional conflicts at a higher abstraction level (e.g., \emph{validity interference}) than reads and writes. Validity interference occurs during a transaction's lifetime when validity assumptions may not hold. 
\ov\ objects are structured hierarchically and may be accessed by all transactions. Conflict detection determines when two transactions cannot both safely commit, to guarantee serializability.

A transaction may read any objects both inside and outside of the validity set, and may write to objects in the invalidity set of the transaction's contract.
There are two cases that transaction interference can occur. The first case is when one transaction presumes the validity of an object that another transaction may invalidate it. It means one transaction's invalidity set overlaps with another's validity set. The second case is when the two invalidity sets overlap.

Transactions in different parts of the object ownership tree may safely execute concurrently. Two transactions whose contracts are $\modelVIn{1}$ and $\modelVIn{2}$ do not interfere with each other if:
$ (\modelV_1 \cap \modelI_2  =  \emptyset)   \wedge  (\modelV_2 \cap \modelI_1  =  \emptyset)  \wedge  (\modelI_1 \cap \modelI_2  =  \emptyset)$.

\paragraph{Contention Management} Validity contracts can be used for conflict resolution to ensure liveness. Contention management was originally proposed for conflict resolution to guarantee progress (livelock avoidance) in obstruction-free STMs \cite{herlihy2003dstm,SXM2005}. Contention management can be tuned to increase throughput dramatically in high contention workloads ~\cite{SCOOL2005,PODC2005}.
With validity contracts, contract-based contention managers can employ different safety-driven strategies to ensure liveness and determined priority of a transaction. For two conflicting transactions, a transaction that is less important or requires significant time for validation, is a more appropriate candidate for abortion.

\paragraph{Validity Check Optimization} Validity checks can be deduced from contracts and scheduled at transaction commits to preserve object validity.
Validation step ensures a transaction never views or makes use of inconsistent data and object revalidation guarantees that transactions only commit valid data. 
Validity checks should not contain any side effects and I/O operations as in ~\cite{Simon2006} and similarly in ours. Validity contracts and subcontracting can be used to reduce the number of validity checks while ensuring safety.

In certain programming scenarios, we may ignore some transaction conflicts
that are less relevant and may in fact be harmless. That is, for an object being read by a transaction and written by another
transaction, a read-write conflict is considered harmless if for example, the read is not covered by the validity assumption (i.e., the transaction's validity
set). For example, generating a booking id in a booking system may not require the validity of the whole system.
Another example is when a client requests for an approximation of the total number of successful transactions.
As a rule of thumbs, if programmers presume the validity of an object for such read access, the contract's validity set must be large enough to cover the read.

\section{Related work}\label{se:related}

This section gives a brief overview of related work in blockchains, smart contracts, and the challenges in programming reasoning.

\subsection{Blockchains and Smart Contracts}

A blockchain is a distributed ledger maintained by peers of a network, all of which follow a consensus protocol.  
It forms a cryptographic chain of blocks, which are replicated among the peers without a single centralized entity. 
State variables are recorded on the blockchain. Transaction, if successful, may change the state of the ledger. A block may contain from zero up to a finite number of transactions. Well known database transaction models such as ACID and BASE are applicable to the blockchain~\cite{tai2017not}.

The term \emph{smart contract} (SC) was first coined by Nick Szabo in 20 years ago. A smart contract is comprised of degitalized promises and protocols within which the parties perform on these promises. A known example of a smart contract is a vending machine; if sufficient money is inserted into the machine, the machine automatically delivers the snack. Smart contracts require specific input parameters and well-defined execution steps, in the form: if `x`, then execute `y`. The actual tasks performed by SCs are deterministic and often trival, such as automatic transfer from one wallet to another, when a certain condition is met.
Latest SC platforms permit trivial to complex instruction sequences in a transaction. 
SCs are issued to trustless system of possibly anonymous parties. If the parties initiate a transaction indicating certain condition is met, the code will execute the triggered step. Each block of transactions is validated to ensure a final consistent ordering. In case transaction is never initiated, the code will not execute. 

Bitcoin Script is the first successful implementation of a blockchain-based smart contract. It is a purposely not-Turing-complete language with a set of simple, pre-defined commands that supports simple forms of SC for Bitcoin transactions. New platforms, such as Ethereum, can support more complex contracts. Solidity, which is a Turing-complete language developed by Ethereum, is one of the most popular languages for SCs.
Newer blockchain platforms, such as Hyperledger Fabric \cite{androulaki2018hyper}, support smart contracts written in various high-level languages.

Conventional smart contract plaforms leverage the \emph{sequential} execution model of transactions within a block. Each block of the blockchain contains multiple smart contract transactions (SCTs). They are executed \emph{sequentially} by the miner before appending the block into the miner's blockchain. On receiving the block, peers as validators will re-execute sequentially the SCTs of the block. If the validators agree with the miner about the final state of the block on their records, the block will be added to the blockchain.

There have been several research works that study concurrent execution of transactions in a block ~\cite{anjana2019efficient,kokoris2018omniledger}. 
One approach is based on object semantics by miner using STM~\cite{anjana2019efficient}. Concurrent miner stores a BlockGraph (BG) structure, capturing the object-conflict relations among the SCTs inside a block. Validators re-execute the same SCTs concurrently using the BG given by miner.  There is another approach that introduced Optimistic Validator to address the double-spending issue by using counter to find any malicious block proposed by a malicious miner~\cite{kokoris2018omniledger}.
In general, SC transactions on accounts in a blockchain can be seen as threads using concurrent objects in shared memory~\cite{sergey2017concur}.
A smart contract transaction can be viewed as a concurrent method, often with some semantic dependencies. When a block commits multiple smart contract transactions simultaneously, it is similar to the way a transactional data structure commits multiple concurrent methods in a single atomic step.

In this paper, we consider the next generation of smart contracts platforms that support concurrent execution of smart contracts in a block. We address the main challenge to ensure the validity of objects in these concurrent smart contracts.

\subsection{Program Reasoning}

Object-oriented programming (OOP) provides basic concepts, such as encapsulation, inheritance and polymorphism, to model softwares with objects showing a clear program structure. 
Decoupling the internal functioning of objects from their external interface can make changes to the internal implementation a lot easier during upgrading or maintenance without affecting the clients. Such practices also enable software reuse and composability.

Object invariants describe essential properties of objects that must be maintained to guarantee the expected behaviour of the program execution. Any violation of the object invariant is often undesirable as it violates program logic and can cause programs to crash.
Yet reasoning about and maintaining object invariants in concurrent programs is difficult. The difficulties come from arbitrary object aliases, mutable object state, methods' side effects and the sophisticated mixture of multiple threads and shared mutable state typically of concurrent O-O programs.

Incorporation of lock-based synchronization primitives, such as monitors, mutexes, and semaphores, in programs may resolve latent defects (data races). Yet they lead to other harder to solve problems such as deadlocks, lock convoying and priority inversion \cite{Herlihy1993}.
At a high level of abstraction, using locks lead to problems like inheritance anomaly, loss of abstraction and code reusability, just to name a few.
In general, these synchronization mechanisms complicate reasoning about and maintenance of concurrent code and object validity.

Exploiting the available concurrency is important to speed up programs \cite{sutter05revoluation}. But concurrent programming is challenging since it requires careful coordination between threads that access shared memory locations.
Pessimistic lock-based synchronization mechanisms, such as locks, semaphores and monitors, provide a means to synchronize memory and coordinate threads, yet the solutions target performance in a cumbersome and error-prone manner.
Lock-based synchronization is difficult to reason about due to: the need to explicitly manage locks and the challenge of composing lock-based code. It can lead to \emph{obvious defects} (such as deadlocks and livelocks), and \emph{latent defects} (such as race conditions), due to the nondeterministic nature of concurrent programs.
Adding more resources, such as faster CPU, disk or network, and more memory, can make a seemingly stable system become unstable; latent defects show up more quickly.

Software Transactional Memory (STM) systems \cite{Shavit1995stm} promise to ease programming concurrent applications by hiding the complexity of concurrency control from programmers and enabling safe composition of scalable applications.
STM offers a programming model based from the ACID properties (e.g. Atomicity, Consistency, Isolation and Durability) of database transactions widely used in software systems.
Atomicity for code wrapped inside an atomic block ensures the code is executed transactionally and if the transaction fails, its effects are discarded. Isolation guarantees that no transaction can observe any partial effects of the others.
Few STMs~\cite{Simon2006} pay attention to data invariants.

This paper presents our new programming language \ov\ to allow developers to write more secure smart contracts that can be statically checked and reasoned about. We also provide a new Solidity API that allows programmers to incorporate validity contracts specifications into smart contracts written in Solidity.

\section{Conclusion}\label{se:con}

Reasoning about the validity of the object states is challenging for concurrent smart contracts, which are to be deployed on  blockchains with trustless parties. 
Despite of being the most popular programming language for smart contracts, Solidity still lacks a facility for specifying and verifying object validity.

To address these shortcomings, we have proposed an approach toward the next generation of smart contract platforms that support optimistic execution of SC transactions.
We have introduced a new Validity API written in Solity to allow developers to specify object invariants and validity contracts in Solidity. We also have presented a new general programming model and a new language, namely \ov, to help developers to write more secure validity-based smart contracts.
The \ov\ language is described with type systems and dynamic sematics. With validity contracts, validity-based smart contracts  are useful for reasoning about the validity of object states in optimistically executed smart contracts and they can be statically checked. In addition, we have shown several possible optimizations for optimistic execution of SC transactions.

\clearpage
\section{Reference}\label{se:ref}

\renewcommand\refname{\vskip -1cm}
\bibliographystyle{abbrv}
\bibliography{citation}

\clearpage
\section{Appendix}\label{se:appendix}

This section gives some more examples of smart contracts, written in Solidity and \ov.

\subsection{Example: Token Contract in Solidity}
\label{apx:tokeninsolid}

\begin{lstlisting}[language=sol]
pragma solidity >=0.4.22 <0.7.0;

library Balances {
	function move(mapping(address => uint256) 
		storage balances, address from, address to, uint amount) internal {
		require(balances[from] >= amount);
		require(balances[to] + amount >= balances[to]);
		balances[from] -= amount;
		balances[to] += amount;
	}
}

contract Token {
	mapping(address => uint256) balances;
	using Balances for *;
	mapping(address => mapping (address => uint256)) allowed;
	
	event Transfer(address from, address to, uint amount);
	event Approval(address owner, address spender, uint amount);
	
	function transfer(address to, uint amount) public returns (bool success) {
		balances.move(msg.sender, to, amount);
		emit Transfer(msg.sender, to, amount);
		return true;	
	}
	
	function transferFrom(address from, address to, uint amount) public returns (bool success) {
		require(allowed[from][msg.sender] >= amount);
		allowed[from][msg.sender] -= amount;
		balances.move(from, to, amount);
		emit Transfer(from, to, amount);
		return true;
	}
	
	function approve(address spender, uint tokens) public returns (bool success) {
		require(allowed[msg.sender][spender] == 0, "");
		allowed[msg.sender][spender] = tokens;
		emit Approval(msg.sender, spender, tokens);
		return true;
	}
	
	function balanceOf(address tokenOwner) public view returns (uint balance) {
		return balances[tokenOwner];
	}
}
\end{lstlisting}

\clearpage
\subsection{Example: Token Contract in \ov}
\label{apx:tokeninov}

\begin{lstlisting}[language=ov]
class Token[o] {
	map<this> balances; // (address => uint256)
	map allowed; // (address => mapping (address => uint256)) 
 	
 	// inv : sum of all balances == N 
 	
	event Transfer(address from, address to, uint amount);
	event Approval(address owner, address spender, uint amount);
	
	void move(address from, address to, uint amount) <this,this> {
		require(balances[from] >= amount);
		require(balances[to] + amount >= balances[to]);
		balances[from] -= amount;
		balances[to] += amount;
	}
	
	public bool transfer(address to, uint amount) <this,this>{
		balances.move(msg.sender, to, amount);
		emit Transfer(msg.sender, to, amount);
		return true;	
	}
	
	public bool transferFrom(address from, address to, uint amount) <this,this> {
		require(allowed[from][msg.sender] >= amount);
		allowed[from][msg.sender] -= amount;
		balances.move(from, to, amount);
		emit Transfer(from, to, amount);
		return true;
	}
	
	public bool approve(address spender, uint tokens) <this,bot> {
		require(allowed[msg.sender][spender] == 0, "");
		allowed[msg.sender][spender] = tokens;
		emit Approval(msg.sender, spender, tokens);
		return true;
	}
	
	public uint balanceOf(address tokenOwner) <this,bot> {
		return balances[tokenOwner];
	}
}
\end{lstlisting}

\clearpage
\subsection{Example: Purchase Contract in Solidity}
\label{apx:purchaseinsolid}

\begin{lstlisting}[language=sol]

pragma solidity ^0.4.22;

contract Purchase {
	uint public value;
	address public seller;
	address public buyer;
	enum State { Created, Locked, Inactive }
	State public state;
	
	// Ensure that `msg.value` is an even number.
	// Division will truncate if it is an odd number.
	// Check via multiplication that it wasn't an odd number.
	constructor() public payable {
		seller = msg.sender;
		value = msg.value / 2;
		require((2 * value) == msg.value, "Value has to be even.");
	}
	
	modifier condition(bool _condition) {require(_condition); _;}	
	modifier onlyBuyer() {require(msg.sender == buyer,"Only buyer can call this."); _;}	
	modifier onlySeller() {require(msg.sender == seller,"Only seller can call this."); _;}	
	modifier inState(State _state) {require(state == _state,"Invalid state."); _;}
	
	event Aborted();
	event PurchaseConfirmed();
	event ItemReceived();
	
	/// Abort the purchase and reclaim the ether.
	/// Can only be called by the seller before
	/// the contract is locked.
	function abort() public onlySeller inState(State.Created) {
		emit Aborted();
		state = State.Inactive;
		seller.transfer(address(this).balance);
	}
	
	/// Confirm the purchase as buyer.
	/// Transaction has to include `2 * value` ether.
	/// The ether will be locked until confirmReceived
	/// is called.
	function confirmPurchase() public inState(State.Created) condition(msg.value == (2 * value))	payable	{
		emit PurchaseConfirmed();
		buyer = msg.sender;
		state = State.Locked;
	}
	
	/// Confirm that you (the buyer) received the item.
	/// This will release the locked ether.
	function confirmReceived()	public onlyBuyer inState(State.Locked) {
		emit ItemReceived();
		// It is important to change the state first because
		// otherwise, the contracts called using `send` below
		// can call in again here.
		state = State.Inactive;
		
		// NOTE: This actually allows both the buyer and the seller to
		// block the refund - the withdraw pattern should be used.		
		buyer.transfer(value);
		seller.transfer(address(this).balance);
	}
}
\end{lstlisting}

\clearpage

\subsection{Example: Purchase Contract in \ov}
\label{apx:purchaseinov}

\begin{lstlisting}[language=ov]
class Purchase[o] {
	public uint value;
	public address<this> seller;
	public address<this> buyer;
	enum State { Created, Locked, Inactive }
	public State<this> state;
	
	// Ensure that `msg.value` is an even number. Division will truncate if it is an odd number.
	// Check via multiplication that it wasn't an odd number.
	public Purchase() {
		seller = msg.sender;
		value = msg.value / 2;
		require((2 * value) == msg.value, "Value has to be even.");
	}
	
	//modifier condition(bool _condition) { require(_condition); _; }
	//modifier onlyBuyer() {	require( msg.sender == buyer, "Only buyer can call this."); _;}
	//modifier onlySeller() { require(msg.sender == seller, "Only seller can call this."); _;}
	//modifier inState(State _state) { require(state == _state, "Invalid state."); _;	}
	event Aborted();
	event PurchaseConfirmed();
	event ItemReceived();
	
	/// Abort the purchase and reclaim the ether. Can only be called by the seller before
	/// the contract is locked.
	/// Pre: onlySeller && inState(State.Created)
	public void abort() <this,this>	{
		emit Aborted();
		state = State.Inactive;
		seller.transfer(address(this).balance);
	}
	
	/// Confirm the purchase as buyer. Transaction has to include `2 * value` ether.
	/// The ether will be locked until confirmReceived is called.
	/// Pre: inState(State.Created) && condition(msg.value == (2 * value))
	public void confirmPurchase() <this,this> {
		emit PurchaseConfirmed();
		buyer = msg.sender;
		state = State.Locked;
	}
	
	/// Confirm that you (the buyer) received the item. This will release the locked ether.
	/// Pre: onlyBuyer && inState(State.Locked)	
	public void confirmReceived() <this,this>{
		emit ItemReceived();
		// Change the state first because otherwise, the contracts called using `send` below
		// can call in again here.
		state = State.Inactive;
		
		// This actually allows both the buyer and the seller to
		// block the refund - the withdraw pattern should be used.
		buyer.transfer(value);
		seller.transfer(address(this).balance);
	}
}
\end{lstlisting}

\clearpage

\subsection{Example: Auction Contract in Solidity}
\label{apx:auctioninsolid}

\begin{lstlisting}[language=sol,multicols=2,basicstyle=\sffamily\tiny]

pragma solidity >0.4.23 <0.5.0;

contract BlindAuction {
	struct Bid {
		bytes32 blindedBid;
		uint deposit;
	}
	
	address public beneficiary;
	uint public biddingEnd;
	uint public revealEnd;
	bool public ended;
	
	mapping(address => Bid[]) public bids;
	
	address public highestBidder;
	uint public highestBid;
	
	// Allowed withdrawals of previous bids
	mapping(address => uint) pendingReturns;
	
	event AuctionEnded(address winner, uint highestBid);
	
	/// Modifiers are a convenient way to validate inputs to
	/// functions. `onlyBefore` is applied to `bid` below:
	/// The new function body is the modifier's body where
	/// `_` is replaced by the old function body.
	modifier onlyBefore(uint _time) { require(now < _time); _; }
	modifier onlyAfter(uint _time) { require(now > _time); _; }
	
	constructor(uint _biddingTime, uint _revealTime, address _beneficiary) public {
		beneficiary = _beneficiary;
		biddingEnd = now + _biddingTime;
		revealEnd = biddingEnd + _revealTime;
	}
	
	/// Place a blinded bid with `_blindedBid` = keccak256(value,
	/// fake, secret).
	/// The sent ether is only refunded if the bid is correctly
	/// revealed in the revealing phase. The bid is valid if the
	/// ether sent together with the bid is at least "value" and
	/// "fake" is not true. Setting "fake" to true and sending
	/// not the exact amount are ways to hide the real bid but
	/// still make the required deposit. The same address can
	/// place multiple bids.
	function bid(bytes32 _blindedBid) public payable onlyBefore(biddingEnd)	{
		bids[msg.sender].push(Bid({
			blindedBid: _blindedBid,
			deposit: msg.value
		}));
	}
	
	/// Reveal your blinded bids. You will get a refund for all
	/// correctly blinded invalid bids and for all bids except for
	/// the totally highest.
	function reveal(uint[] _values, bool[] _fake, bytes32[] _secret) 
		public onlyAfter(biddingEnd) onlyBefore(revealEnd)	{
		uint length = bids[msg.sender].length;
		require(_values.length == length);
		require(_fake.length == length);
		require(_secret.length == length);
		
		uint refund;
		for (uint i = 0; i < length; i++) {
		Bid storage bid = bids[msg.sender][i];
		(uint value, bool fake, bytes32 secret) =
		(_values[i], _fake[i], _secret[i]);
		if (bid.blindedBid != keccak256(value, fake, secret)) {
			// Bid was not actually revealed.
			// Do not refund deposit.
			continue;
		}
		refund += bid.deposit;
		if (!fake && bid.deposit >= value) {
			if (placeBid(msg.sender, value))
			 refund -= value;
			}
			// Make it impossible for the sender to re-claim
			// the same deposit.
			bid.blindedBid = bytes32(0);
		}
		msg.sender.transfer(refund);
	}
	
	// This is an "internal" function which means that it
	// can only be called from the contract itself (or from
	// derived contracts).
	function placeBid(address bidder, uint value) internal	returns (bool success)
	{
		if (value <= highestBid) {
			return false;
		}
		if (highestBidder != 0) {
			// Refund the previously highest bidder.
			pendingReturns[highestBidder] += highestBid;
		}
		highestBid = value;
		highestBidder = bidder;
		return true;
	}
	
	/// Withdraw a bid that was overbid.
	function withdraw() public {
		uint amount = pendingReturns[msg.sender];
		if (amount > 0) {
			// It is important to set this to zero because the recipient
			// can call this function again as part of the receiving call
			// before `transfer` returns (see the remark above about
			// conditions -> effects -> interaction).
			pendingReturns[msg.sender] = 0;
			
			msg.sender.transfer(amount);
		}
	}
	
	/// End the auction and send the highest bid
	/// to the beneficiary.
	function auctionEnd() public onlyAfter(revealEnd) {
		require(!ended);
		emit AuctionEnded(highestBidder, highestBid);
		ended = true;
		beneficiary.transfer(highestBid);
	}
}

\end{lstlisting}

\clearpage

\subsection{Example: Auction Contract in \ov}
\label{apx:auctioninov}

\begin{lstlisting}[language=ov,multicols=2,basicstyle=\scriptsize]

class BlindAuction[o] {
	class Bid[o] {
		bytes32 blindedBid;
		uint deposit;
	}
	
	public address<this> beneficiary;
	public bool<this> ended;
	
	public map<this> bids; // (address => Bid[])
	
	public address<this> highestBidder;
	public uint<this> highestBid;
	
	// Allowed withdrawals of previous bids
	map<this> pendingReturns; // (address => uint)
	
	event AuctionEnded(address winner, uint highestBid);
	
	BlindAuction(uint _biddingTime, 
			uint _revealTime, address _beneficiary) {
		beneficiary = _beneficiary;
	}
	
	/// Place a blinded bid
	public void bid(bytes32 _blindedBid) <this,this> {
		bids[msg.sender].push(Bid({
			blindedBid: _blindedBid,
			deposit: msg.value
		}));
	}
	
	/// Reveal your blinded bids
	public void reveal(	uint[] _values, 
			bool[] _fake, bytes32[] _secret ) <this,this> {
		uint length = bids[msg.sender].length;
		require(_values.length == length);
		require(_fake.length == length);
		require(_secret.length == length);
		
		uint refund;
		for (uint i = 0; i < length; i++) {
			Bid storage bid = bids[msg.sender][i];
			(uint value, bool fake, bytes32 secret) =
			(_values[i], _fake[i], _secret[i]);
			if (bid.blindedBid != keccak256(value, fake, secret)) {
				// Bid was not actually revealed.
				// Do not refund deposit.
				continue;
			}
			refund += bid.deposit;
			if (!fake && bid.deposit >= value) {
				if (placeBid(msg.sender, value))
					refund -= value;
			}
			// Make it impossible for the sender to re-claim
			// the same deposit.
			bid.blindedBid = bytes32(0);
		}
		msg.sender.transfer(refund);
	}
	
	// place a bid
	private bool placeBid(address bidder, uint value) <this,this> {
		if (value <= highestBid) {
			return false;
		}
		if (highestBidder != 0) {
			// Refund the previously highest bidder.
			pendingReturns[highestBidder] += highestBid;
		}
		highestBid = value;
		highestBidder = bidder;
		return true;
	}
	
	/// Withdraw a bid that was overbid.
	public void withdraw() <this,this> {
		uint amount = pendingReturns[msg.sender];
		if (amount > 0) {
			// It is important to set this to zero because 
			// the recipient can call this function again
			pendingReturns[msg.sender] = 0;
			
			msg.sender.transfer(amount);
		}
	}
	
	/// End the auction and send the highest bid 
	/// to the beneficiary.
	public void auctionEnd() <this, this> {
		require(!ended);
		emit AuctionEnded(highestBidder, highestBid);
		ended = true;
		beneficiary.transfer(highestBid);
	}
}

\end{lstlisting}

\clearpage

\subsection{Example: Ballot contract in Solidity}
\label{apx:ballotinsolid}

\begin{lstlisting}[language=sol,multicols=2,basicstyle=\sffamily\tiny]
pragma solidity ^0.4.0;
/// @title Voting with delegation.
contract Ballot {
	// This declares a new complex type which will
	// be used for variables later.
	// It will represent a single voter.
	struct Voter {
		uint weight; // weight is accumulated by delegation
		bool voted;  // if true, that person already voted
		address delegate; // person delegated to
		uint vote;   // index of the voted proposal
	}
	// This is a type for a single proposal.
	struct Proposal
	{
		bytes32 name;   // short name (up to 32 bytes)
		uint voteCount; // number of accumulated votes
	}
	address public chairperson;
	// This declares a state variable that
	// stores a `Voter` struct for each possible address.
	mapping(address => Voter) public voters;
	// A dynamically-sized array of `Proposal` structs.
	Proposal[] public proposals;
	/// Create a new ballot to choose one of `proposalNames`.
	function Ballot(bytes32[] proposalNames) {
		chairperson = msg.sender;
		voters[chairperson].weight = 1;
		// For each of the provided proposal names,
		// create a new proposal object and add it
		// to the end of the array.
		for (uint i = 0; i < proposalNames.length; i++) {
			// `Proposal({...})` creates a temporary
			// Proposal object and `proposals.push(...)`
			// appends it to the end of `proposals`.
			proposals.push(Proposal({
				name: proposalNames[i],
				voteCount: 0
			}));
		}
	}
	// Give `voter` the right to vote on this ballot.
	// May only be called by `chairperson`.
	function giveRightToVote(address voter) {
		if (msg.sender != chairperson || voters[voter].voted) {
		// `throw` terminates and reverts all changes to
		// the state and to Ether balances. It is often
		// a good idea to use this if functions are
		// called incorrectly. But watch out, this
		// will also consume all provided gas.
			throw;
		}
		voters[voter].weight = 1;
	}
	/// Delegate your vote to the voter `to`.
	function delegate(address to) {
		// assigns reference
		Voter sender = voters[msg.sender];
		if (sender.voted)
			throw;
		// Forward the delegation as long as
		// `to` also delegated.
		// In general, such loops are very dangerous,
		// because if they run too long, they might
		// need more gas than is available in a block.
		// In this case, the delegation will not be executed,
		// but in other situations, such loops might
		// cause a contract to get "stuck" completely.
		while (
			voters[to].delegate != address(0) &&
			voters[to].delegate != msg.sender
		) {
			to = voters[to].delegate;
		}
		// We found a loop in the delegation, not allowed.
		if (to == msg.sender) {
			throw;
		}
		// Since `sender` is a reference, this
		// modifies `voters[msg.sender].voted`
		sender.voted = true;
		sender.delegate = to;
		Voter delegate = voters[to];
		if (delegate.voted) {
			// If the delegate already voted,
			// directly add to the number of votes
			proposals[delegate.vote].voteCount += sender.weight;
		} else {
			// If the delegate did not vote yet,
			// add to her weight.
			delegate.weight += sender.weight;
		}
	}
	/// Give your vote (including votes delegated to you)
	/// to proposal `proposals[proposal].name`.
	function vote(uint proposal) {
		Voter sender = voters[msg.sender];
		if (sender.voted)
			throw;
		sender.voted = true;
		sender.vote = proposal;
		// If `proposal` is out of the range of the array,
		// this will throw automatically and revert all
		// changes.
		proposals[proposal].voteCount += sender.weight;
	}
	/// @dev Computes the winning proposal taking all
	/// previous votes into account.
	function winningProposal() constant
		returns (uint winningProposal)
	{
		uint winningVoteCount = 0;
		for (uint p = 0; p < proposals.length; p++) {
			if (proposals[p].voteCount > winningVoteCount) {
				winningVoteCount = proposals[p].voteCount;
				winningProposal = p;
			}
		}
	}
	// Calls winningProposal() function to get the index
	// of the winner contained in the proposals array and then
	// returns the name of the winner
	function winnerName() constant
		returns (bytes32 winnerName)
	{
		winnerName = proposals[winningProposal()].name;
	}
}
\end{lstlisting}

\newpage
\subsection{Example: Ballot contract in \ov}
\label{apx:ballotinov}

\begin{lstlisting}[language=ov, multicols=2, basicstyle=\scriptsize]

// @title Voting with delegation.
class Ballot[o] {
	// a single voter.
	class Voter[oo] {
		uint weight; // weight is accumulated by delegation
		bool voted;  // if true, that person already voted
		address delegate; // person delegated to
		uint vote;   // index of the voted proposal
		
		// inv: !voted || weight > 0 
	}
	// This is a type for a single proposal.
	class Proposal[oo] {
		bytes32 name;   // short name (up to 32 bytes)
		uint voteCount; // number of accumulated votes
	}
	public address<this> chairperson;
	public map<this> voters; // (address => Voter)
	public Proposal[]<this> proposals;
	
	/// Create a new ballot to choose one of `proposalNames`.
	Ballot(bytes32[] proposalNames) <this,this> {
		chairperson = msg.sender;
		voters[chairperson].weight = 1;
		for (uint i = 0; i < proposalNames.length; i++) {
			proposals.push(Proposal({
				name: proposalNames[i],
				voteCount: 0
			}));
		}
	}	
	// Give `voter` the right to vote on this ballot.
	// May only be called by `chairperson`.
	void giveRightToVote(address voter) <this,bot> {
		if (msg.sender != chairperson || voters[voter].voted) {
			// `throw` terminates and reverts all changes to
			// the state and to Ether balances.
			throw;
		}
		voters[voter].weight = 1;
	}
	/// Delegate your vote to the voter `to`.
	void delegate(address to) <this,bot> {
		// assigns reference
		Voter sender = voters[msg.sender];
		if (sender.voted)
			throw;
		// Forward the delegation as long as `to` also delegated.
		while (
			voters[to].delegate != address(0) &&
			voters[to].delegate != msg.sender
		) {
			to = voters[to].delegate;
		}
		// We found a loop in the delegation, not allowed.
		if (to == msg.sender) {
			throw;
		}
		// Since `sender` is a reference, this
		// modifies `voters[msg.sender].voted`
		sender.voted = true;
		sender.delegate = to;
		Voter delegate = voters[to];
		if (delegate.voted) {
			// If the delegate already voted,
			// directly add to the number of votes
			proposals[delegate.vote].voteCount += sender.weight;
		} else {
			// If the delegate did not vote yet,
			// add to her weight.
			delegate.weight += sender.weight;
		}
	}
	/// Give your vote (including votes delegated to you)
	/// to proposal `proposals[proposal].name`.
	void vote(uint proposal) <this,this> {
		Voter sender = voters[msg.sender];
		if (sender.voted)
			throw;
		sender.voted = true;
		sender.vote = proposal;
		// If `proposal` is out of the range of the array, 
		// this will throw automatically and revert all changes.
		proposals[proposal].voteCount += sender.weight;
	}
	/// Computes the winning proposal taking all previous votes into account.
	uint winningProposal() <this,bot> {
		uint winningProposal;
		uint winningVoteCount = 0;
		for (uint p = 0; p < proposals.length; p++) {
			if (proposals[p].voteCount > winningVoteCount) {
				winningVoteCount = proposals[p].voteCount;
				winningProposal = p;
			}
		}
		return winningProposal;
	}
	
	/// returns the name of the winner
	bytes32 winnerName() <this,bot>	{
		bytes32 winnerName = proposals[winningProposal()].name;
		return winnerName
	}
}
\end{lstlisting}

\clearpage

\end{document}